\documentclass[9pt,twocolumn,twoside]{osajnl}
\journal{ol} % Choose journal (ao,jocn,josaa,josab,ol,optica,pr)

\setboolean{shortarticle}{true}
\usepackage{graphicx}
\title{Rapid Scan White Light Pump-Probe Spectroscopy with 100 kHz Shot-to-Shot Detection}

\author[1,2]{Vivek N. Bhat}
\author[1,2]{Asha S. Thomas}
\author[1]{Vivek Tiwari*}

\affil[1]{Solid State and Structural Chemistry Unit, Indian Institute of Science, Bengaluru, KA 560012, India}
\affil[2]{equal authors}

\affil[*]{Corresponding author:vivektiwari@iisc.ac.in}

\begin{abstract}
We demonstrate a femtosecond pump-probe spectrometer which utilizes a white light supercontinuum as input, and relies on mutual synchronization of acousto-optical chopper, pump-probe delay stage and the CCD camera to record shot-to-shot pump-probe spectra while the pump-probe delay is scanned synchronously with the laser repetition rate. The unique combination of technologies implemented here allows for electronically controllable and  repetition-rate scalable detection throughput that is only limited by the camera frame rate. Despite RMS white-light probe fluctuations of $\sim$5.5\%, fully leveraging the temporal correlations in white light and fine sampling of pump-probe delay along with 30x reduction in equivalent data collection time compared to stepwise scanning leads to reduction of RMS noise without multichannel referencing down to $\sim$0.33 mOD for a scattering nanotube sample. This demonstration opens door for impulsive pump-probe micro-spectroscopy of scattering samples with broadband spectral coverage and minimized sample exposure.
\end{abstract}

\setboolean{displaycopyright}{true}

\begin{document}
\maketitle
%[motivate TA usefulness]
Ultrafast transient absorption (TA) spectroscopy has been instrumental in enabling mechanistic insights\cite{Cerullo2020review,Kukura2015} into condensed phase ultrafast phenomenon ranging from femtosecond charge transfer reactions in solution, energy transfer in excitonically coupled systems such as photosynthetic proteins, and ultrafast exciton diffusion and dissociation in photovoltaic thin films. Although more sophisticated multidimensional spectroscopies\cite{OgilvieARPC,Tiwari2021} are available, the power of TA lies in its relative ease of instrumentation and data processing. TA approaches have also been extended to nonlinear imaging\cite{Warren2016Review,Grumstrup2019} where the additional spatial information has allowed to disentangle the effects of sample morphology on electronic relaxation.

%[motivate white light pump, need for multichannel detection, associated complexity, shot-to-shot as a way to bypass it, end with motivating rapidScan]
While several TA approaches use a narrowband pump and broadband supercontinuum probe, approaches based on a broadband pump are particularly insightful\cite{Kukura2015} for inferring structural dynamics accompanying femtosecond internal conversion between electronic states.  Although broadband pump is typically generated through noncollinear optical parameteric amplification (NOPA) for reasons of significantly higher RMS stability\cite{Lang2018,Bradler2009a} across the entire bandwidth, white light continuum (WLC) pump is relatively much easier to generate and therefore desirable to implement in TA. However, higher RMS fluctuations in WLC along with spectral and temporal correlations\cite{Bradler2014} necessitate either longer averaging times, or introducing multichannel referencing and  balanced detection\cite{Ernsting2010,Bradler2014,Turner2015,Lang2018} to achieve desired signal-to-noise (SNR) levels. Alternatively, a few approaches\cite{Lang2018} have implemented higher averaging through high-repetition rate shot-to-shot detection, which additionally leverages correlations between consecutive laser shots to suppress laser noise without the complexity of multichannel reference detection. For example, Brixner and co-workers \cite{Brixner2014} have demonstrated 100 kHz shot-to-shot detection with stepwise  scan of pump-probe delay $T$. It is also known\cite{Moon1993,Bartels2012} that rapid scanning of $T$ delay can achieve higher averaging through reduction in single scan time. However, in case of WLC sources, it may not by itself be effective in suppressing the low-frequency component of 1/$f$ laser noise, which manifests as spectral and temporal correlations\cite{Bradler2014} in WLC. So far, TA approaches that combine the advantages of rapid scanning with shot-to-shot detection are scarce\cite{Lang2018}, with limited capability in terms of repetition rate and $T$ scan range. 
%[what is new in present work]
Here we combine the advantages of a broadband WLC source, high-repetition rate shot-to-shot detection and rapid $T$ scanning to demonstrate $\sim$30x reduction in data collection time and noise floor down to $0.33$ mOD in femtosecond pump-probe spectra of a scattering nanotube sample. The detection throughput of our approach is electronically tunable and only limited by the camera frame rate, while $T$ delay is sampled densely by scanning synchronously with the laser repetition rate. The particular combination of technologies implemented here makes our approach repetition-rate scalable and especially suitable for high throughput, impulsive pump-probe microscopy with efficient scatter suppression, without added cost and complexity of light sources, multichannel detection or long sample exposure.
%A broadband pump not only provides high time resolution but also impulsively excites coherent vibronic superpositions whose dephasing dynamics encodes vital parameters regarding the role of bath in dictating system trajectory. Such dynamics can also be influenced by local morphology and thus it is desirable to be able to access the same information in TA microscopy approaches as well.

\begin{figure}[h!]
	\centering
	\fbox{\includegraphics[width=3 in]{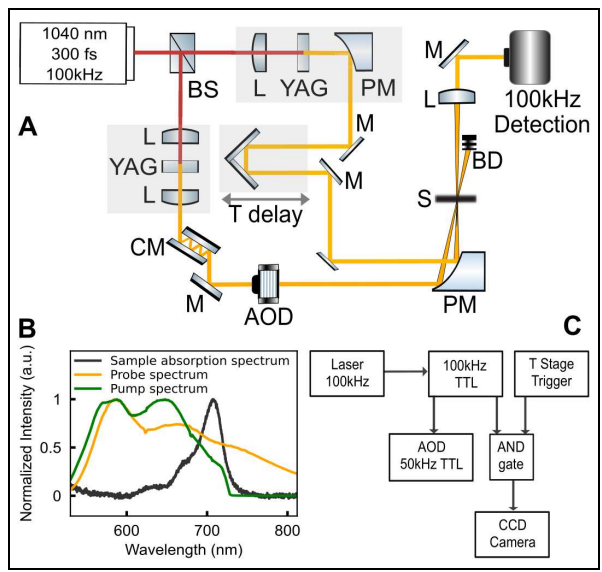}}
	\caption{(A) Experimental schematic of the rapid scan white light pump-probe spectrometer. BS, beam splitter; L, lens; CM, chirp mirror; M, mirror; AOD, acousto optic deflector; PM, Parabolic mirror; S, Sample; BD, beam dump. 100KHz detection denotes the spectrograph, line camera and timing circuit than enables shot-to-shot detection. (B) Sample absorption spectrum overlaid with pump and probe spectra derived from the WLC. (C) Electronics synchronization scheme.}
	\label{fig:fig1}
\end{figure}

The experimental setup schematic is shown in Figure \ref{fig:fig1} (panel A). Fundamental beam of 1040 nm from a Yb:KGW amplifier (Spirit One) at 100 kHz is split into two arms of $\sim$1 $\mu$J pulse energy to generate pump and probe WLC lines. The pump and probe continua are filtered using a 725 nm and 850 nm shortpass filters, respectively. The pump arm is routed through a pair of chirped mirrors for dispersion pre-compensation, followed by an acousto-optical deflector (AOD).

\begin{figure}[h!]
	\centering
	\fbox{\includegraphics[width=3 in]{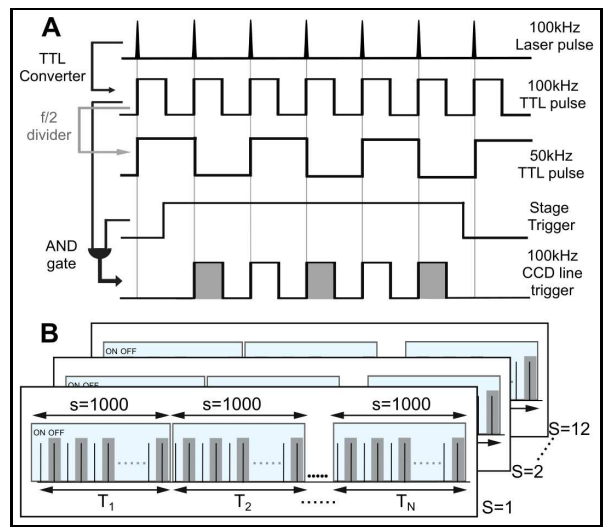}}
	\caption{(A) Timing diagram showing the synchronization of laser repetition rate to AOD, CCD camera and stage trigger. Shaded region in AND gate output represents the pump OFF state. (B) Shot-to-shot data acquisition and averaging scheme. $S$ denotes a given $T$ scan, $s$ denotes the number of probe spectra binned together for one $T$ step, with total $T_1$ to $T_N$ binned time steps. In the data presented in Fig.\ref{fig:fig4}, $S$ =12 and $s$=1000.}
	\label{fig:fig2}
\end{figure}

The AOD is synced to the laser repetition rate $f$ through a $f/2m$ TTL divider such that every other pump pulse is blocked from exciting the sample when $m=1$. The probe arm is directly routed to the sample without dispersion pre-compensation. The pump and probe arms are focused into a sample flowcell (pathlength 200 $\mu$m) using reflective optics. The sample is circulated at a rate of $\sim$70 ml/min using a peristaltic pump to ensure a fresh sample spot every 10$\mu$s. After the sample, the pump beam is blocked while the probe is routed to the spectrograph using a combination of reflective optics and achromatic lens. Every probe shot is recorded by a line CCD camera (e2v AViiVA, 14×28 $\mu$m 1024 pixels) attached to the spectrograph. Panel B shows the average pump and probe laser spectrum measured at the sample location, and overlaid with the sample absorption spectrum. \\  
Fig.~\ref{fig:fig1}C and Fig.~\ref{fig:fig2}A show the timing electronics which is the crucial component of the experiment. The $f = $ 100kHz laser pulse train is converted to a TTL pulse train, which is then split into two signals. One part is converted to a $f/2$ TTL signal used to drive the RF controller of the AOD pump chopper. The second part is send into a AND circuit along with a trigger signal from the $T$ stage controller. The stage controller is set to output a high signal whenever the stage is within a defined $T$ delay range, such that the AND gate triggers the CCD camera only within this defined range. The camera reads every probe shot, with readout rate only limited by the camera line rate. Note that $m=2$ or larger does not correspond to shot-to-shot detection but is electronically controllable in the above implementation, if so desired. 

%This is illustrated in Table 1.\\
%
%\begin{table}[htbp]
%\centering
%\caption{\bf Tunability of data collection schemes}
%\begin{tabular}{ccc}
%\hline
%Frequency division  &   frequency 
%&  number of continuous \\ f = laser repetition rate & & pump on or off shots(m)\\
%\hline
%$f/2$ & $50$kHz & $1$(Shot-to-Shot) \\
%$f/4$ & $25$kHz & $2$ \\
%$f/8$ & $12.5$kHz & $4$ \\
%\hline
%\end{tabular}
%\end{table}\label{table1}

%Figure \ref{fig:fig2} describes the data collection scheme implemented here. %\vt{See Brixner\cite{Brixner2014}, Zanni\cite{Kearns2017} and Cohen\cite{Cohen2017} for example. Describe how each velocity corresponds to a given time-period, and how each time-step corresponds to a given number of frames. Use inline equations. Describe stage velocity profile, encoder details. See Lang2018 Section IIIA for example. Mention clearly the data collection time. Mention that certain overlapping regions are maintained for assessing and minimizing effects of stage velocity on spot alignment in the line array during post processing. Then refer to this in the more detailed discussion in Fig 4.}

The shot-to-shot data collection scheme is presented in Fig.~\ref{fig:fig2}B. With rapid $T$ scan, stage velocity $v$ and laser repetition rate $f$ determine $T$ time step given by $(2m/f).(2v/c)$. For the case of shot-to-shot detection ($m=1$), consecutive probe shots with and without pump (gray shaded pulses) determine a time step and the corresponding pump-probe spectrum. The total $T$ delay in Fig.~\ref{fig:fig4} ranges from -0.75 ps to 26.85 ps. This range is divided into 3 intervals with uniform velocities 0.075mm/s, 0.15mm/s and 0.75mm/s, with velocities increasing for larger $T$ intervals. The delay stage (ILS150BPP,Newport) is moved continuously using the SGamma motion profile of Newport XPS motion controller which eliminates the sudden change in acceleration in a trapezoidal motion profile. The distance to be covered to accelerate to the constant velocity and decelerate to the stop is added before and after the trigger window respectively so that stage moves with uniform velocity within the trigger window during data acquisition. 
In each interval, $432$ frames, each with $1000$ probe shots are collected so that the sample is continuously exposed for only $\sim$4.4 seconds before a mechanical shutter turns on to block the pump and probe beams. For shot-to-shot detection with $s$ consecutive probe shots, total $s-1$ nearest neighbor pump ON-OFF pairs can be averaged together to form a binned time step $T_i$ and the corresponding averaged pump-probe signal.  

The data is collected in three intervals as $-0.75$ to $1.45$ ps in $5$ fs binned steps, $0.95$ to $5.35$ ps in $10$ fs binned steps, and $4.85$ to $26.85$ ps in $50$ fs binned steps. In each case, finely sampled $T$ steps are 500x smaller. $N=1232$ is the total number of binned $T$ delay points after processing. $T$ scan is repeated $S=12$ times to break the low-frequency 1/$f$ correlations in the WLC \cite{Bradler2014}. The time taken for single scan in each interval is 4.4 seconds, hence the total experimental time is 2.64 minutes \textit{including} $S=12$ scans. Note that binning $s-1$ finely sampled $T$ steps together is what provides effective suppression of high-frequency scatter, as demonstrated\cite{Dahlberg2017} by Engel and co-workers in the context of \textit{in vivo} multidimensional spectroscopy. In comparison, for a stepwise pump-probe scan, a nominal wait time of 300 ms at each coarse time step already leads to a $\sim$30x slower experiment. 

%The bin sizes in the three intervals are 5 fs, 10 fs and 50 fs, respectively. It is calculated as bin-size(fs) $=$ $\frac{2\times velocity \times repetition\hspace{1mm}period\times s}{c}$. Factor 2 is to account for the probe travelling twice the distance in the retro reflector path, c is the speed of light in $\mu$m/fs, repetition period = 10.18$\mu$s in our case, velocity in mm/s.  

\begin{figure}[h!]
	\centering\includegraphics{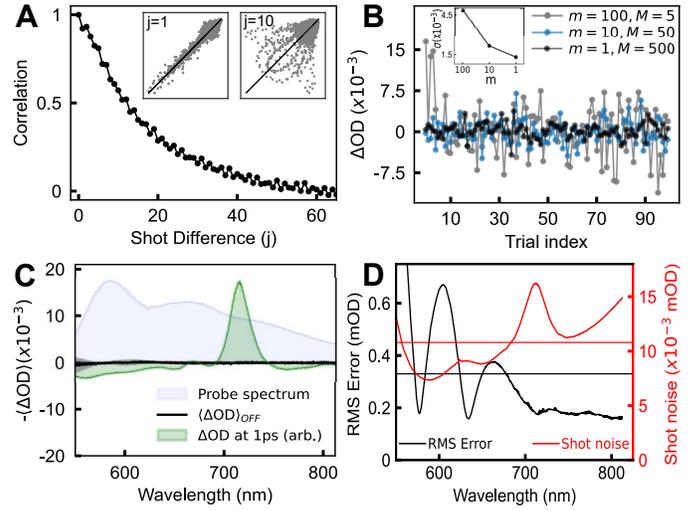}
	\caption{(A) Laser intensity autocorrelation as a function of shot separation $j$ for 10$^5$ consecutive probe shots transmitted through the sample with pump blocked, insets: scatter maps for $j=1$ and $j=10$. (B) $\Delta$OD calculated with data of panel A assuming data data collection schemes. $m$ denotes number of probe shot with pump ON and OFF that lead to one data point. $m=1$ corresponds to shot-to-shot detection and corresponds to least standard deviation (inset) form the expected zero signal. (C) Average of $\Delta$OD with error bar calculated with the experimental data collection scheme of Fig.~\ref{fig:fig2}B but with pump blocked. Green shows $\Delta$OD signal at $T$ = 1 ps (with pump ON) on the same scale. Averaged probe spectrum is overlaid for reference. (D) Comparison of RMS Error in $\Delta$OD with the shot noise limit. Horizontal lines indicates the averaged values from 550nm to 812nm region.}
	\label{fig:fig3} 
\end{figure}

\begin{figure}[h!]
	\centering\includegraphics{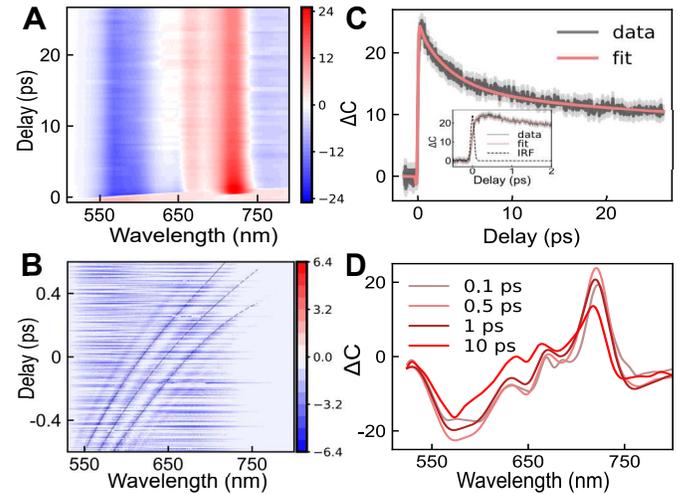}
	\caption{(A) 2D contour plot of $\Delta$C pump-probe spectra of porphyrin nanotubes. (B) 2D contour plot of the chirp correction signal with the third-order polynomial fit overlaid. (C) Difference signal decay at 720nm overlaid with the error bar and the global fit of the decay. The Gaussian IRF of 95fs FWHM is shown in the inset. (D) Chirp-corrected pump-probe spectra at fixed $T$.}
	\label{fig:fig4}
\end{figure}

%\vt{This is the most important part, Figure 2 was the second most important part. Read Leonard et al. Fig. 3 \cite{Leonard2016} for generating the expected shot-noise limited graph. On the same we can plot the RMS vs wavelength as well. Also make a table similar to Brixner for comparing noise levels. We want to be able to show that the final $\mu$OD levels of noise is better than (?) Brixner, how does it compare to Lang2018 Table 1. Use equations to explain the analysis. your description should highlight the points that 1. correlation exists until i to i+nth shot but that itself is successively reduced, the strongest correlation is always between consecutive pairs, $N-1$ of those. Correlation is useful to maintain along $T$ dynamics, but no so useful if a converged average value at a given $T$ was desired, see Lang2018 Section III for this point. 2. No global structures due to spectral and temporal correlations are seen in the averaged "dark" spectra with pump blocked and no sample, see Riedle \cite{Bradler2014} Fig 1 about this point. 3. the last point is about the correlations which decay as seen in panel a and panel d.}
Figure \ref{fig:fig3} presents the noise analysis of the experiment. Following prior analysis\cite{Lang2018}, Fig.~\ref{fig:fig3}A computes the autocorrelation between 10$^5$ consecutive probe shots after transmission through the nanotube sample with pump blocked as a function of shot separation $j$. The closer shots (smaller j) are more correlated. This loss of correlation is better seen as scatter plots (in the insets) for $j=1$ versus $j=10$. Panel A suggests strong correlations between consecutive WLC laser shots can be leveraged to bypass multichannel referencing and balanced detection. This is further confirmed in panel B which uses the 10$^5$ probe shots in panel A to calculate\cite{Brixner2014} $\Delta OD =\langle{-\log_{10}\frac{C_{ON}}{C_{OFF}}}\rangle$ for shot-to-shot ($m=1$) and non shot-to-shot ($m=10,100$) schemes. As defined earlier, $m$ is the number of probe shots (with pump ON or OFF) that are averaged together to calculate a difference spectrum, that is, pump-probe spectrum. $M$ is the number of difference spectra recorded for a fixed $m$. The standard deviation of $\Delta OD$ signal (inset), around the expected mean value of zero, is $1.33\times10^{-3}$ for shot-to-shot scheme, which is $\sim$1.6 and $\sim$3.6 times lesser than the non shot-to-shot schemes ($m=10, 100$ respectively), suggesting that the advantages of shot-to-shot detection\cite{Brixner2014} are maintained even in the presence of a scattering sample.\\

%, as -- 
%\begin{equation} 
%    \rho_{ij} = \frac{\displaystyle\sum_{i=1} ^{i=N-j}[E_i - \overline{E}] [E_{i+j} - \overline{E}]}{\displaystyle\sum_{i=1} ^{i=N}[E_i - \overline{E}]^2},
%\end{equation}
%where $E_i$ is the spectrally integrated energy of $i^{th}$ shot and $\overline{E} $ is the average of spectrally integrated pulses. \vt{is notation same or different from Cerullo ?} 

Fig.~\ref{fig:fig3}C calculates the average shot-to-shot $\Delta$OD signal for $(s, S) = (1000,12)$, considering $s-1$ pairs for averaging with sample present and pump pulse blocked, such that a zero signal is expected. This averaging is the same as that employed for the pump-probe data in Fig.~\ref{fig:fig4}. The resulting $\Delta$OD curve is nearly flat (within the error bar) over the entire probe bandwidth suggesting effective averaging of spectral and temporal WLC correlations\cite{Bradler2014}. The measurement error (gray shaded) ranges from $\sim$ 0.21-0.64 mOD. The average RMS error of 0.33 mOD is calculated as $\Delta OD_{RMSE}=\frac{1}{\log(10)}.\frac{\sqrt{\overline{(I_i-I_{i+1})^2}}}{\overline{I}}$, where $I_i$ corresponds to $i^{th}$ probe shot and bar denotes average over all possible pairs. $\Delta$OD signal at $T$ = 1 ps with pump unblocked is included for comparison. A scaled probe spectrum is also included for reference. Fig.~\ref{fig:fig3}D compares this RMS error with the expected shot noise limited calculated as $\Delta$OD$_{SN}$ $= \frac{1}{\log(10)}.\frac{1}{\sqrt{N_{Ph}}}$ where $N_{Ph}$ is the number of photons falling on the CCD sensor. $N_{Ph}$ can be calculated as $\frac{\text{Pixel Intensity (Counts)}}{\text{max  Intensity (for 12bit)}}.\frac{\text{Full Well Capacity}}{\text{Quantum Efficieny}}$, with average quantum efficiency of 0.68 in the probe spectral region, 12 bit resolution and full well capacity of $2.38\times10^5$.  The individual noise contributions summarized in Table \ref{table2} show that $\sim$50x difference between $\Delta OD_{RMSE}$ and the shot-noise limit arises from the dominant laser noise in the WLC.
%\begin{table}[h!]
%\centering
%\small
%\caption{\bf Noise and the errors in the experiments.}
%\resizebox{\columnwidth}{!}{%
%\begin{tabular}{cccc}
%\hline
% & Electronic Noise  &  Shot noise &  Laser intensity fluctuation\\
%\hline
%value & $\overline{C} =$ 3408 counts& $N_{Ph}=2.91\times10^5$ & $\overline{C} =$ 3408 counts \\
%error & 7.7 counts& $\sqrt{N_{Ph}}=539.6$ & 315 counts \\
%relative error & $2.26\times10^{-3}$ & $\frac{1}{\sqrt{N_{Ph}}} = 1.8\times10^{-3}$ & 0.0924\\
%Absolute error& $2.063\times10^{-5}$& $1.69\times10^{-5}$ & $0.844\times10^{-3}$\\
%\hline
%\end{tabular}
%}
%\end{table}  \label{table2}
\begin{table}[h!]
	\centering
	\small
	\caption{\bf Comparison of noise contributions.}
	\resizebox{\columnwidth}{!}{%
		\begin{tabular}{cccc}
			\hline
			& Electronic Noise  &  Shot noise &  Laser intensity fluctuation\\
			\hline
			mean value & $\overline{C} =$ 3408 counts& $N_{Ph}=2.91\times10^5$ & $\overline{C} =$ 3408 counts \\
			noise & 7.7 counts& $\sqrt{N_{Ph}}=539.6$ & 187 counts \\
			relative error & $2.3\times10^{-3}$ & $\frac{1}{\sqrt{N_{Ph}}} = 1.8\times10^{-3}$ & $54.8\times10^{-3}$\\
%			Absolute error& $2.063\times10^{-5}$& $1.69\times10^{-5}$ & $0.844\times10^{-3}$\\
			\hline
		\end{tabular}
	}
\end{table}  \label{table2}
\normalsize

%Figure \ref{fig:fig4} describes the experimental results.  
%\vt{Describe chirp correction procedure we follow for Figure \ref{fig:fig4} panel B. Describe the analysis in panel c. Mention that the movement of spot on the camera as the stage is scanned is crucial to check in order to ensure no artifacts arise due to misalignment in the line array, mention how we checked it. Mention also the checks about movement of focal spot with $T$ and refer to a SI figure. Refer to overlapping regions being averaged and the averaged spectra are multiplied by a average of constant factor to take care of \% misalignments (refer back to discussion of Fig 2). After corrections mention how do the \% misalignments compare to noise level. }

%\vt{Finally describe Figure \ref{fig:fig4} panel d. relate it to the timescales known in literature and comment in 2-3 lines about consistency(?) with literature and photophysics being probed. Mention clearly that this is a sample that scatters a lot and special double chopping techniques are employed in order to suppress scatter in 2DES, cite Zigmantas here \cite{Zigmantas2011}. Mention that error bars on the SRPP curves are of the order of XY $\mu$OD and include sample scatter contributions apart from those described in Figure 3.}
Using the above described approach, Fig.~\ref{fig:fig4} presents femtosecond pump-probe measurements on a porphyrin nanotube sample. Following previous protocols\cite{Huang2014}, 1mM meso-Tetra(4-sulfonatophenyl)porphine (TPP) dihydrochloride (Santa Cruz Biotechnology) in Ethanol was added to equal volume of 0.07N HCl/EtOH. Steady state absorption spectrum (Fig.~\ref{fig:fig1}B) confirmed the formation of nanotubes. The pump and probe focal spot sizes at the sample location were measured to be $\sim$25$\mu$m and $\sim$29$\mu$m ($1/e^2$), amounting to a pump and probe fluence of 150 and 250$\mu$J/cm$^2$, respectively. Over this range of fluence, the measured \% transmission of both pump and probe, as well as the pump-probe signal, was confirmed to be linear with pump and probe power. The exciton density was calculated\cite{Kriete2019a} to be 1 per 116 molecules which is lower than the annihilation threshold for the Q band excitons which are delocalized over 88 molecules (when B-Q coupling is neglected)\cite{Huang2014}.\\

Fig.~\ref{fig:fig4}A is the measured contour map of the pump-probe signal $\Delta$C without probe chirp correction. Although analysis in Fig.~\ref{fig:fig3} considers $\Delta$OD to be consistent with prior reports, we consider changes in probe photon number ($\Delta$C) because it directly relates\cite{Cho2013} to the pump-probe signal even if changes in absorption coefficient over the sample pathlength are not small. During setup alignment prior to measurements, any trends in the probe focal spot due to $T$ stage movement were ruled out using a CMOS camera at the focal spot while the stage was continuously moved. Further, probe alignment on the CCD line array was cross verified by scanning the stage in the three scan intervals with the pump blocked. Compared to 5.5\% RMS probe fluctuations (Table \ref{table2}), maximum deviation in the averaged probe spectra due to $T$ scanning was only 1.9\% of the total counts at that pixel.  Further, overlapping scan range (of 0.5 ps) were chosen to ensure an overlapping region of $\sim$0.5 ps in order to measure and correct for any residual minor deviations during data processing. The data collected in the overlapping regions (denoted by $R=1,2,3$) are compared in terms of the mean over the overlapping region, denoted by $\langle{{\Delta}{C}}\rangle_R$. Minor deviations between different $R$ are then easily corrected by a factor given by the ratio between the overlapping regions. In the measured data, after ensuring the above alignment checks, deviations between $\langle{{\Delta}{C}}\rangle_R$ were found to be of the order of $\Delta$C error bar and easily corrected using the ratio. Chirp correction function is determined by collecting pump scatter and probe interference fringes from an aggregated anthracene solution as a function of $T$, and fitting the resulting fringes to a third order polynomial (Fig.~\ref{fig:fig4}B). The coefficients of this polynomial serve as the initial input for dispersion curve optimization in the GloTarAn\cite{glotaran}. The decay of signal maxima (720 nm) after above corrections is shown in Fig.~\ref{fig:fig4}C. A 95 fs IRF suggests significant higher-order dispersion not compensated by chirped mirrors, but may be effectively compensated through liquid crystal or deformable mirror based pulse compressors. It does not however affect the central theme of this report. Pump-probe spectra for representative $T$ points are shown in Fig.~\ref{fig:fig4}D. The 2D dataset is interpolated to a uniform delay axis to avoid any loss of signal during chirp correction due to the larger delay steps at later $T$. From the interpolated data, difference spectra at each $T$ have been sliced along a polynomial corresponding to the dispersion curve, followed by Fourier filtering to eliminate high-frequency noise arising due to interpolation. \\

%The spike in the 0.1ps trace at 668 and 685nm, where the signal is weak, corresponds to the non-resonant coherent artifact contribution to the signal which is expected for the setup with an IRF of 95fs. \\
%The pump-probe scatter interference fringes, prominent near zero pump-probe delays are recorded as a function of the waiting time $T$. For each of the fringes collected, the midpoint of the fringe at a specific T can be traced to measure how the $T$ is dispersed over the probe spectral range ($T(\lambda)$). 

%The deviation among the probe spectra recorded at fixed T delays and while the stage is in continuous motion at the three different velocities used in the scan are within the limit of the probe white light fluctuations (15.48\% and 15.04\%, of WLC fluctuation respectively). 
%A $1280\times720$ pixel CMOS camera was kept at the focal point, and the spot was recorded while the T stage was in continuous motion. The spot profile was analyzed in ImageJ and MATLAB for the spot size and maxima position[ refer SI]. The ratio $r=\frac{ \langle{{\Delta}{C_1}}\rangle}{{\langle{{\Delta}{C_2}}}\rangle}$ ) between regions 1,2 is 1 irrespective of the wavelength in an ideal scenario. However due to any residual misalignment if there is a deviation of this ratio from 1, the latter interval is scaled by a factor which is the mean of r over the probe wavelength range. 

Global fitting of the data reveals a tri-exponential decay with $\sim$371.5 fs, $\sim$3.57 ps, and $\sim$95.8 ps decay constants. The faster timescales correspond\cite{Kano2002} to intra-Q band electronic relaxation and vibrational relaxation, respectively. The longest decay component of 95.8ps can be attributed to the Q-band exciton lifetime although a $\sim$27 ps scan is insufficient to describe it accurately. The positive $\Delta{C}$ signal at 720nm and 670nm correspond to the $Q_x$ and $Q_y$ bleach signals. The broad excited state absorption (ESA) at 573nm corresponds to transitions to higher excitons while the red-shifted ESA at 765nm corresponds to transitions from the $k\neq 1$ exciton states to multi-excitonic states\cite{Kano2002}. The above spectral features and dynamics on TPP nanotubes are in good agreement to the reported literature, with low error bars despite the sample being highly scattering and in continuous flow. While sophisticated double chopping techniques\cite{Zigmantas2011} were implemented to suppress the scatter from TPP nanotubes, it should be emphasized that the above data results from a combination of 100 kHz shot-to-shot detection and fine $T$ sampling through rapid scan, without employing any additional scattering suppression techniques. The approach demonstrated here provides a promising path towards high-throughput white-light pump-probe microscopy of scatter prone samples with minimized sample exposure.

\begin{backmatter}
\bmsection{Funding} Content in the funding section will be generated entirely from details submitted to Prism. Authors may add placeholder text in the manuscript to assess length, but any text added to this section in the manuscript will be replaced during production and will display official funder names along with any grant numbers provided. If additional details about a funder are required, they may be added to the Acknowledgments, even if this duplicates information in the funding section. See the example below in Acknowledgements.

\bmsection{Acknowledgments} VNB acknowledges research fellowship from DST-Inspire. AST acknowledges Prime Minister's Research Fellowship, MoE India. VT acknowledges the Infosys Young Investigator Fellowship supported by the Infosys Foundation, Bangalore. 

\bmsection{Disclosures} The authors declare no conflicts of interest.

\bmsection{Data availability} Data underlying the results presented in this paper are not publicly available at this time but may be obtained from the authors upon reasonable request.

%\bmsection{Supplemental document}
%See Supplement 1 for supporting content. 

\end{backmatter}

%\section{References}
%
%Note that \emph{Optics Letters} and \emph{Optica} short articles use an abbreviated reference style. Citations to journal articles should omit the article title and final page number; this abbreviated reference style is produced automatically when the \emph{Optics Letters} journal option is selected in the template, if you are using a .bib file for your references.
%
%However, full references (to aid the editor and reviewers) must be included as well on a fifth informational page that will not count against page length; again this will be produced automatically if you are using a .bib file.
%
%\bigskip
%\noindent Add citations manually or use BibTeX. See \cite{Zhang:14,OSA,FORSTER2007,testthesis,manga_rao_single_2007}.

% Bibliography
\bibliography{rapidSpp}

% Full bibliography added automatically for Optics Letters submissions; the following line will simply be ignored if submitting to other journals.
% Note that this extra page will not count against page length
\bibliographyfullrefs{rapidSpp}

%Manual citation list
%\begin{thebibliography}{1}
%\bibitem{Zhang:14}
%Y.~Zhang, S.~Qiao, L.~Sun, Q.~W. Shi, W.~Huang, %L.~Li, and Z.~Yang,
 % \enquote{Photoinduced active terahertz metamaterials with nanostructured
  %vanadium dioxide film deposited by sol-gel method,} Opt. Express \textbf{22},
  %11070--11078 (2014).
%\end{thebibliography}

% Please include bios and photos of all authors for aop articles
\ifthenelse{\equal{\journalref}{aop}}{%
\section*{Author Biographies}
\begingroup
\setlength\intextsep{0pt}
\begin{minipage}[t][6.3cm][t]{1.0\textwidth} % Adjust height [6.3cm] as required for separation of bio photos.
  \begin{wrapfigure}{L}{0.25\textwidth}
    \includegraphics[width=0.25\textwidth]{john_smith.eps}
  \end{wrapfigure}
  \noindent
  {\bfseries John Smith} received his BSc (Mathematics) in 2000 from The University of Maryland. His research interests include lasers and optics.
\end{minipage}
\begin{minipage}{1.0\textwidth}
  \begin{wrapfigure}{L}{0.25\textwidth}
    \includegraphics[width=0.25\textwidth]{alice_smith.eps}
  \end{wrapfigure}
  \noindent
  {\bfseries Alice Smith} also received her BSc (Mathematics) in 2000 from The University of Maryland. Her research interests also include lasers and optics.
\end{minipage}
\endgroup
}{}

%\begin{table}[htbp]
%\centering
%\caption{\bf \hsapce{}Noise and the errors in the experiments.}
%\begin{tabular}{cccc}
%\hline
% & Electronic Noise  &  Shot noise &  Laser intensity 
% &  &  &  &fluctuation\\
%\hline
%value & $\overline{C} =$ 3408 counts& $N_{Ph}=2.91\times10^5$ & %$\overline{C} =$ 3408 counts \\
%error & 7.7 counts& $\sqrt{N_{Ph}}=539.6$ & 315 counts \\
%relative error & $2.26\times10^{-3}$ & $\frac{1}{\sqrt{N_{Ph}}} = %1.8\times10^{-3}$ & 0.0924\\
%Absolute error& $2.063\times10^{-5}$& $1.69\times10^{-5}$ & %$0.844\times10^{-3}$\\
%\hline
%\end{tabular}
%  \label{$f$ = repetition rate }
%\end{table}

\end{document}